\documentclass[aps,prb,twocolumn,showpacs]{revtex4}

\usepackage{amsmath,amssymb,amsfonts,upgreek,bm}
\usepackage{color}
\usepackage{graphicx}

\graphicspath{{./}}

\newcommand{\dprime}{'\!'}
\newcommand{\tprime}{'\!'\!'}

\begin{document}

\title{Spin-caloric transport properties of cobalt nanostructures:\\
  spin disorder effects from first principles}

\date{\today}

\author{Roman Kov\'a\v{c}ik}
\email{r.kovacik@fz-juelich.de}
\affiliation{Peter Gr\"unberg Institut and Institute for Advanced
  Simulation, Forschungszentrum J\"ulich and JARA, 52425 J\"ulich, Germany}
\author{Phivos Mavropoulos}
\affiliation{Peter Gr\"unberg Institut and Institute for Advanced
  Simulation, Forschungszentrum J\"ulich and JARA, 52425 J\"ulich, Germany}
\author{Daniel Wortmann}
\affiliation{Peter Gr\"unberg Institut and Institute for Advanced
  Simulation, Forschungszentrum J\"ulich and JARA, 52425 J\"ulich, Germany}
\author{Stefan Bl\"ugel}
\affiliation{Peter Gr\"unberg Institut and Institute for Advanced
  Simulation, Forschungszentrum J\"ulich and JARA, 52425 J\"ulich, Germany}

\begin{abstract}
  The fundamental aspects of spin-dependent transport processes and
  their interplay with temperature gradients, as given by the spin
  Seebeck coefficient, are still largely unexplored and a multitude of
  contributing factors must be considered.  We used density functional
  theory together with a Monte-Carlo-based statistical method to
  simulate simple nanostructures, such as Co nanowires and films
  embedded in a Cu host or in vacuum, and investigated the influence
  of spin-disorder scattering on electron transport at elevated
  temperatures. While we show that the spin-dependent scattering of
  electrons due to temperature induced disorder of the local magnetic
  moments contributes significantly to the resistance, thermoelectric
  and spin-caloric transport coefficients, we also conclude that the
  actual magnitude of these effects cannot be predicted,
  quantitatively or qualitatively, without such detailed calculations.
\end{abstract}

\pacs{75.76.+j, 72.15.Jf, 72.10.-d}

\maketitle

\section{Introduction}\label{sec:intro}

The recent discovery of the spin Seebeck
effect\cite{2008-10//uchida/takahashi//saitoh} triggered a broad
discussion about its microscopic origin and correct interpretation of
the measured data (spin Seebeck versus spin-dependent Seebeck
effect),\cite{2010-03//lu/zhu//zhang,2010-03//tulapurkar/suzuki,2010-11//jaworski/yang//myers,2010-11//slachter/bakker//van-wees,2010-12//adachi/uchida//maekawa,2011-03//adachi/ohe//maekawa,2011-05//jaworski/yang//heremans,2012-03//flipse/bakker//van-wees,2012-07//dejene/flipse/van-wees}
and extended the field of
spin-caloritronics,\cite{2012-05//bauer/saitoh/van-wees} which
investigates the coupling between electrical, spin, and heat
transport. Possible technological importance was suggested for the
spin Seebeck effect in thermally-driven position sensing.
\cite{2011-11//uchida/kirihara//saitoh,2012-03//weiler/althammer//goennenwein}
Furthermore, an enhanced Peltier effect reported in submicron-sized
metallic junctions could lead to applications in electronics
micro-cooling.\cite{2005-01//fukushima/yagami//yuasa,2007-09//katayama-yoshida/fukushima//sato,2010-06//sugihara/kodzuka//fukushima,2011-01//vu/sato/katayama-yoshida}
Ferromagnetic materials subject to a temperature gradient experience,
in addition to a heat current (thermal conductivity) and a charge
current (thermoelectric effect), an induced spin current (spin-caloric
effect). In metals we expect a large part of the spin current to arise
from spin-polarized electron propagation due to generally different
conductivity for the majority and minority spin channels, while other
possible effects, e.g., spin transport due to magnons or even phonons,
are dominant in insulators.

Theoretical and computational investigation is an essential part of
understanding spin-caloric phenomena due to a non-trivial connection
between the microscopic character of relevant materials and their
functionality. The asymmetry of the electronic transmission
coefficient as a function of energy around the Fermi level
($E_\text{F}$) enters the expressions for the thermoelectric effects,
making quantitative or even semi-quantitative predictions next to
impossible on a simple model level. Given the complexity of the
electronic structure, numerical calculations are therefore inevitable.

An important contribution to the electron-transport phenomena at high
temperatures is the formation of a spin-disordered state due to
local-moment fluctuations in the magnetic material. This comes on top
of the phonon contribution at high temperatures and the spin-orbit
contribution to spin mixing that occurs at all
temperatures.\cite{2013-09//popescu/kratzer} The fluctuations induce
spin-conserving and spin-flip scattering and clearly contribute to the
temperature-dependent transport phenomena. For example, they induce
the well-known spin disorder resistivity that has been experimentally
investigated in the past in ferromagnetic
materials,\cite{1964-09//arajs/colvin,1967-07//kierspe/kohlhaas/gonska}
and was successfully modeled with ab-initio
techniques.\cite{2007-05//wysocki/belashchenko//van-schilfgaarde,2008-02//buruzs/szunyogh/weinberger,2009-12//wysocki/sabirianov//belashchenko,2012-06//glasbrenner/belashchenko//turek,2012-10//kudrnovsky/drchal//belashchenko}
It is obvious that spin fluctuations must also contribute to
thermoelectric and spin-caloric phenomena. However, this effect has
not been investigated so far. In the present paper we address this
problematics by means of density-functional calculations and
Monte-Carlo simulations and arrive at the conclusion that the impact
of temperature-induced spin disorder is strong at temperatures where
the magnetization reduction is significant. We also find that there is
no universal correlation between the temperature-dependent
magnetization and the thermoelectric and spin-caloric coupling, i.e.,
it has to be examined separately for each material and microscopic
structure and at each temperature, due to the delicate modulation of
the electron-scattering as a function of energy around $E_\text{F}$
and as a function of temperature.

We focus on magnetic nanostructures, particularly on Co nanocolumns
embedded between Cu leads, motivated by the miniaturization of
spintronics devices and by recent suggestions that nanostructured
magnetic materials (e.g., in the ``Konbu'' phase) can lead to
extraordinary thermoelectric effects due to quantum
confinement.\cite{2010-06//sugihara/kodzuka//fukushima,2011-01//vu/sato/katayama-yoshida}
In addition, we approach the bulk limit by considering the transport
through a thin layer formed by several atomic layers of Co. The
electronic structure of the studied systems is calculated within the
multiple scattering screened Korringa-Kohn-Rostoker Green function
(KKR-GF) framework using the full-potential
formalism.\cite{2002-03//papanikolaou/zeller/dederichs} The
Monte-Carlo methodology is then used to simulate the effect of
temperature on the magnetic configurations within a Heisenberg model
with the exchange coupling parameters calculated according to
Liechtenstein {\it et
al}.\cite{1987-05//liechtenstein/katsnelson//gubanov} The transmission
probability through the spin-disordered magnetic structures is
obtained using the Landauer-B\"uttiker approach for the ballistic
transport within the KKR-GF
framework,\cite{2004-03//mavropoulos/papanikolaou/dederichs} extended
in this work to account for the non-collinear magnetic effects similar
to work in Ref.~\onlinecite{2006-11//yavorsky/mertig}, thus providing
individual spin-preserving and spin-flip contributions to the
transmission probability. The electrical conductance and Seebeck
coefficients are finally calculated from the transport coefficients.

In the following, we describe our model systems and briefly summarize
methodologies used in this work and corresponding computational
details (Sec.~\ref{sec:meth}). Results are presented and discussed in
Sec.~\ref{sec:res} and our main conclusions are given in
Sec.~\ref{sec:sum}.

\section{Methodology}\label{sec:meth}

\subsection{Geometric structure}

The geometry that we choose is intended to serve as a generic model of
Co nanostructures embedded, on the one hand, between
free-electron-metal leads, and on the other hand surrounded in the
lateral direction either by a free-electron metal or by vacuum or an
insulating material modeled here by vacuum. We choose a few different
geometries to see if the effect of spin-disorder on the transport
coefficients in the nano-scale is significantly affected by the
magnetic nanostructure shape and size and by the surrounding medium
(metallic or insulating).

\begin{figure}
  \center
  \includegraphics[width=\columnwidth]{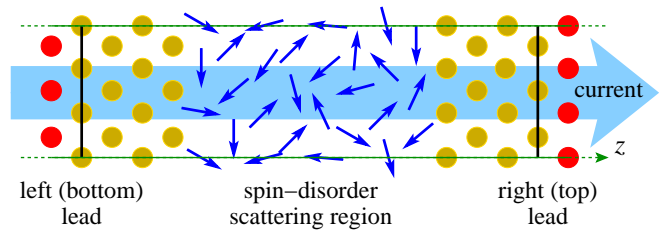}
  \caption{(Color online) General setup of our model system with atoms
    placed on the fcc crystal lattice. The direction of current is
    parallel with the $z$ axis ([001] direction). Cu and Co atoms are
    shown as spheres and arrows, respectively. The transmission
    probability is evaluated for pairs of atomic layers (indicated by
    vertical lines) placed far enough from the Cu/Co interface.}
  \label{fig:setup}
\end{figure}

The basic setup of our model systems is depicted in
Fig.~\ref{fig:setup}. The left and right semi-infinite leads consist
of the fcc crystal lattice of Cu atoms with the experimental lattice
constant $a_\text{lat}=3.62$~\AA\ and with the interface to the
scattering region orthogonal to the $z$ axis ([001] direction). The
region between the leads contains 8 atomic layers of either Co atoms
forming a thin layer or Co atoms in a shape of a thin wire in various
structural configurations surrounded by Cu or vacuum (see
Fig.~\ref{fig:geom}). The sites in the scattering region epitaxially
follow the perfect fcc lattice of the leads, which is an acceptable
approximation due to the small size of this region. Since our focus is
to study the effect of spin disorder, the atoms are kept at their
ideal unrelaxed positions. A supercell approach with two-dimensional
periodicity within the $xy$ plane is used to model the real space spin
disorder in a thin layer of Co atoms [Fig.~\ref{fig:geom}(c)], as well
as for all wire-like Co structures [Fig.~\ref{fig:geom}(a-b,d-f)],
imposing a separation of at least $2a_\text{lat}$ to any atom in their
periodic images. The transmission probability between the left and
right lead is evaluated for pairs of Cu atomic layers placed
sufficiently away from the interface with the magnetic region so that
the Cu potentials are bulk-like.

\subsection{Electronic structure}\label{ssec:met-es}

The KKR-GF method using the full-potential
formalism\cite{2002-03//papanikolaou/zeller/dederichs,1990-09//stefanou/akai/zeller,1991-09//stefanou/zeller,SPR-TB-KKR}
and local density approximation\cite{1980-08//vosko/wilk/nusair} to
the exchange-correlation energy functional is employed to calculate
the electronic structure of our model systems, for which the angular
moment expansion is truncated after $l_\text{max}=3$. We calculate the
electronic structure of the ground state in a multiple step procedure.

In the first step we calculate the self-consistent density and
potential of a ``slab'' system Cu(7)-\{Cu/Co/Va\}(8)-Cu(7) with one
atom per layer. Here, the ``scattering region'' \{Cu/Co/Va\}(8)
consists of eight layers of Co atoms, Cu atoms or vacuum, depending on
the system. This region is sandwiched between two Cu(7) regions,
consisting of seven atomic layers of Cu, and the whole slab is
embedded in vacuum. A well converged density was reached using a
$36{\times}36$ k-point mesh for the integration in the surface
Brillouin zone (SBZ) and a smearing temperature of 800~K.

In a second step, we replace the outermost part of the slab, i.e., the
three outermost Cu atomic layers and the outer vacuum, by
half-infinite Cu leads. Then, employing the decimation
technique,\cite{1986-02//garcia-moliner/velasco,1994-01//szunyogh/ujfalussy//kollar}
we use the central potential of the Cu(7) layer that is already
bulk-like to a very good approximation to construct the self-energy
induced by the half-infinite region. In other words, we attach the
half-infinite leads ($\cdots$Cu- and -Cu$\cdots$) on the
Cu(4)-\{Cu/Co/Va\}(8)-Cu(4) ``central'' part of the slab.

\begin{figure}
  \center
  \includegraphics[width=\columnwidth]{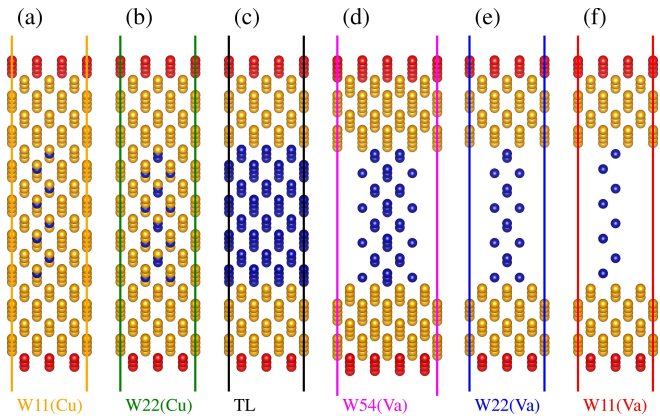}
  \caption{(Color online) Cobalt (blue/dark spheres) nanostructures
    sandwiched between Cu leads (yellow and red/bright spheres) used
    in this study: (c) thin Co layer ``TL'' consisting of 8
    monolayers, (a,f) monoatomic wires ``W11(Cu)'' and ``W11(Va)''
    having 1 Co atom in all layers, (b,e) biatomic wires ``W22(Cu)''
    and ``W22(Va)''having 2 Co atoms in all layers, (d) wire
    ``W54(Va)'' with alternating 5 and 4 Co atoms. The (Cu) and (Va)
    indicate the type of embedding (copper and vacuum, respectively)
    of the Co nanostructure. Periodic boundaries are indicated by
    solid lines. Crystal structures were plotted using
    VESTA.\cite{2011-12//momma/izumi}}
  \label{fig:geom}
\end{figure}

\begin{figure}
  \center
  \includegraphics[width=\columnwidth]{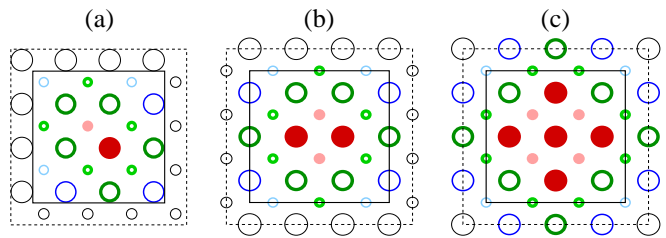}
  \caption{(Color online) Setup of the nanowires geometry for the (a)
    W11, (b) W22 and (c) W54 nanowires: Co sites (red / filled
    circle), their first nearest neighbor (green / thick line circle),
    second nearest neighbor (blue / medium thick line circle) and
    embedding sites (black / thin line circle) are shown in the two
    adjacent layers (large and small circles, respectively). The
    $3{\times}3$ and $4{\times}4$ supercell is indicated by solid and
    dashed line, respectively.}
  \label{fig:impurity}
\end{figure}

The third and final step corresponds to the construction of supercell
potentials. At this point, the $\cdots$Cu-Cu(4)-Co(8)-Cu(4)-Cu$\cdots$
system forms a base for the model system of the thin Co layer (TL). A
$3{\times}3$ supercell [shown in Fig.~\ref{fig:geom}(c)] is
constructed by replicating the individual site potentials. To obtain
the electronic structure of all other systems (nanowires), the
embedded Co wire and the nearest neighbors were further treated
self-consistently by the impurity Green function
method.\cite{bauer-thesis,2002-03//papanikolaou/zeller/dederichs}
Here, we use as a reference the Cu(4)-Cu(8)-Cu(4) system (Cu bulk) for
the nanowires embedded in Cu [Fig.~\ref{fig:geom}(a,b)] or the
Cu(4)-Va(8)-Cu(4) system for the nanowires embedded in vacuum
[Fig.~\ref{fig:geom}(d,e,f)]. The structure of the nanowires is shown
from a side view in Fig.~\ref{fig:geom}(a-b,d-f) and from a top view
in Fig.~\ref{fig:impurity}. Inclusion of the second nearest neighbors
in the impurity cluster led to negligible differences of occupation
($< 0.006$~electron) and magnetic moments ($< 0.005\ \mu_{\text{B}}$)
of the cobalt atoms and their nearest neighbor sites. The occupation
of second nearest neighbors did not differ more than 0.011 electron
from the unperturbed reference site in their respective
layer. Finally, a supercell in the $xy$ plane, i.e., perpendicular to
the lead/wire interface was formed by the converged potentials of the
impurity cluster sites embedded in the respective reference potentials
of \{Cu/Va\}. The nanowires W11(Cu/Va) and W22(Cu/Va) were modeled in
the $3{\times}3$ supercell, whereas the W54(Va) wire, having a larger
cross-section, was modeled in the $4{\times}4$ supercell (see also
Fig.~\ref{fig:geom}).

\subsection{Spin disorder}\label{ssec:met-sd}

Our model of the spin disorder is based on adopting the moment
directions as they are given by a classical Heisenberg model at
non-zero temperature. In this spirit we set up a Heisenberg
Hamiltonian
\begin{equation}\label{eq:hh}
  H=-\sum_{ij}J_{ij}\ \mathbf{M}_i\cdot\mathbf{M}_j
  ,
\end{equation}
where $\mathbf{M}_i$ and $\mathbf{M}_j$ are unit vectors pointing in
the direction of the moments at sites $i$ and $j$, respectively, while
$J_{ij}$ are the exchange parameters extracted from the ground-state
electronic structure. Based on the formalism of Liechtenstein {\it et
  al.},\cite{1987-05//liechtenstein/katsnelson//gubanov} we calculate
the exchange coupling parameters $J_{ij}$ between the Co atoms.

The thermal fluctuations of the magnetic moments of Co atoms are
modeled by the Monte-Carlo (MC) approach using the Metropolis
algorithm.\cite{1953-06//metropolis/rosenbluth//teller} As a random
number generator we used the Mersenne
twister.\cite{1998-01//matsumoto/nishimura} Since the statistics of
quantities evaluated from the spin configurations strongly depends on
the temperature and character of a particular system, we empirically
determine the number of required MC configurations $N_{\text{conf}}$
by a simple criterion that accounts for the fluctuation amplitude,
\begin{equation}\label{eq:nconf}
  N_{\text{conf}}=N_0
  \sqrt{\langle{m^2_{\text{MC}}}\rangle-\langle{m_{\text{MC}}}\rangle^2}
  /\langle{m_{0}}\rangle
  ,
\end{equation}
where $m_{\text{MC}}$ is the magnitude of the system magnetization at
a given MC snapshot, $\langle{m_{\text{MC}}}\rangle$ its average over
MC configurations, $\langle{m^2_{\text{MC}}}\rangle$ the MC average
of $m_{\text{MC}}^2$ and $\langle{m_{0}}\rangle$ the magnitude of the
ground-state magnetization in the KKR-GF calculation. The empirical
constant $N_0$ was set to 5000 yielding typical $N_\text{conf}\approx
800$ around the crossover temperature where the fluctuations are most
pronounced. The spin-up and spin-down directions in the electronic
structure derived quantities at each MC snapshot are given with
respect to the global magnetization axis of the same MC snapshot and
averaged at the end over all snapshots. Besides monitoring the
evolution of average magnetic moment and magnetic susceptibility as a
function of temperature, we calculate the correlation function $C_N$
between moments at different layer pairs, where $Na_\text{lat}/2$ is
the distance between the two layers. In the calculation of $C_N$ we
include correlations between moments at atomic sites whose distance, when projected
onto the $xy$ plane, does not exceed $0.5\ a_\text{lat}$. Using
indices $a$ and $b$ for the in-plane position and $c$, $d$ for the
layers, we define
\begin{equation}\label{eq:zcorr}
  C_{N}(T)=
  \frac{1}{N_aN_b}
  \sum_{a,b}
  \frac{1}{N_{cd}}
  \sum_{c < d}\langle{\mathbf{M}_{abc}\cdot\mathbf{M}_{abd}}\rangle_T
  ,
\end{equation}
where $N_{cd}$ is the number of all layer pairs with distance
$Na_\text{lat}/2$, and $N_a$ and $N_b$ is the number of magnetic sites
along the $x$ and $y$ directions of the MC supercell.

\subsection{Electron transport}\label{ssec:met-tran}

In order to evaluate the transmission probability matrix using the
formalism of non-collinear magnetism, a code was developed interfacing
the existing KKR-GF\cite{2002-03//papanikolaou/zeller/dederichs} and
transport\cite{2004-03//mavropoulos/papanikolaou/dederichs} programs,
which were modified accordingly to treat the non-collinear magnetic
states. We make implicit use of the adiabatic approximation, assuming
that the electron traverses the nanostructure or junction at a faster
timescale compared to the precession of localized
moments.\cite{1996-07//antropov/katsnelson//kusnezov,1998-07//halilov/eschrig//oppeneer}
Our approach amounts to a rotation of the ground-state magnetic part
of the site-dependent potentials in the direction prescribed by the MC
snapshot. This is done without a further self-consistent calculation
of the non-collinear state, as such a calculation requires many
self-consistent steps while adjusting the necessary constraining
fields, leading to an increase of computational time by one to two
orders of magnitude. The resulting absolute values of magnetic moments
differ only marginally from their respective values at the ground
state, confirming the dominance of the intra-atomic exchange
interaction over the inter-atomic ones. In the following, we briefly
outline our implementation of the calculation of transport properties
within the non-collinear formalism.

In the spirit of multiple-scattering theory and the KKR-GF method, we
apply the spin rotations on the site-dependent $t$-matrices that are
then used to calculate the non-collinear Green function via the Dyson
equation. Thus the $t$-matrices are calculated as spin-diagonal
quantities in a local spin frame (indicated by $t^{\mu(\text{loc})}$)
and they are transformed to the global frame where they are indicated
by $t^{\mu(\text{glob})}$.\cite{2005-12//lounis/mavropoulos//blugel}
The transformation matrix $\mathbf{U}^{\mu}$, corresponding to the
standard spherical rotation angles $\theta^{\mu}$ and $\phi^{\mu}$ at
the site $\mu$, given by
\begin{equation}\label{eq:rotmat}
  \mathbf{U}^{\mu}=
  \Bigg[
  \begin{matrix}\
    \cos(\theta^{\mu}/2)\ \text{e}^{-\frac{\text{i}}{2}\phi^{\mu}} &
    -\sin(\theta^{\mu}/2)\ \text{e}^{-\frac{\text{i}}{2}\phi^{\mu}} \\[2pt]
    \sin(\theta^{\mu}/2)\ \text{e}^{\frac{\text{i}}{2}\phi^{\mu}} &
    \cos(\theta^{\mu}/2)\ \text{e}^{\frac{\text{i}}{2}\phi^{\mu}}
  \ \end{matrix}
  \Bigg]
  ,
\end{equation}
is used to mix the spin-up and spin-down components of the local
$t$-matrix at energy $\varepsilon$, resulting in a $2{\times}2$ matrix
in spin space ($\sigma=\,\uparrow,\downarrow$)
\begin{equation}\label{eq:tglob}
  \mathbf{t}_{LL'}^{\mu(\text{glob})}(\varepsilon)=
  \mathbf{U}^{\mu}
  \,
  \text{diag}\left[
  t_{LL'}^{\mu\uparrow(\text{loc})}(\varepsilon),
  t_{LL'}^{\mu\downarrow(\text{loc})}(\varepsilon)
  \right]
  (\mathbf{U}^{\mu})^{\dagger}
  ,
\end{equation}
where $L=(l,m)$ is the angular momentum quantum number and
$t_{LL'}^{\mu\sigma(\text{loc})}(\varepsilon)$ is calculated from the
spin-dependent potential $V^{\mu\sigma}$ as
\begin{equation}\label{eq:tloc}
  t_{LL'}^{\mu\sigma(\text{loc})}(\varepsilon)=
  \int{
    \text{d}\mathbf{r} \, j_{L}(\mathbf{r},\varepsilon) 
    V^{\mu\sigma}(\mathbf{r}) \,
    R_{L'}^{\mu\sigma}(\mathbf{r},\varepsilon)
  }
  ,
\end{equation}
where $j_L$ and $R_L^{\mu\sigma}$ are the Bessel functions and
standard radial solutions to the Kohn-Sham potential at site $\mu$,
respectively. The structural Green function matrix $\bm{\mathcal{G}}$
of the system with dimension
$2{\times}N_\text{at}{\times}(l_\text{max}+1)^2$, with $N_\text{at}$
the number of atoms, is obtained via the Dyson equation
\begin{equation}\label{eq:gfstr}
  \bm{\mathcal{G}}(\varepsilon)=
  \bm{\mathcal{G}}_0(\varepsilon)
  \left\{
  \mathbf{I}-
  \big[
  \mathbf{t}^{(\text{glob})}(\varepsilon)-\mathbf{t}_0(\varepsilon)
  \big]^{-1}
  \bm{\mathcal{G}}_0(\varepsilon)
  \right\}^{-1}
  ,
\end{equation}
where $\mathbf{I}$ is the unit matrix and $\bm{\mathcal{G}}_0$ and
$\mathbf{t}_0$ are the reference system Green function matrix and
$t$-matrix, respectively. After a Fourier transform, taking the
periodic supercell geometry into account, the Green function is
calculated for each momentum channel $\mathbf{k}_{\parallel}$. The
transmission probability matrix in spin space as a function of
$\mathbf{k}_{\parallel}$ and $\varepsilon$ is calculated as
\begin{eqnarray}\label{eq:condkres}
  \Gamma^{\sigma\sigma'}(\varepsilon)=&
  \displaystyle
  \sum_{\mu\mu'}
  \sum_{LL'}
  \sum_{L\dprime L\tprime}
  (J_{LL\dprime}^{\mu\sigma}-J_{L\dprime L}^{\mu\sigma*})
  (J_{L'L\tprime}^{\mu'\sigma'}-J_{L\tprime L'}^{\mu'\sigma'*})
  \nonumber\\
  &\times \,
  \mathcal{G}_{LL'}^{\mu\mu'\sigma\sigma'}
  \,
  \mathcal{G}_{L\dprime L\tprime}^{\mu\mu'\sigma\sigma'*}
  ,
\end{eqnarray}
where $J_{LL'}^{\mu\sigma}$ is the corresponding current-density
matrix element in the non-magnetic lead in a cell associated with the
site $\mu$ with the volume $\Omega_\mu$
\begin{equation}\label{eq:jll}
  J_{LL'}^{\mu\sigma}(\varepsilon)=
  \frac{1}{d_\text{at}}\int_{\Omega_\mu} \text{d}\mathbf{r} \,
  R_{L}^{\mu\sigma}(\mathbf{r},\varepsilon) \,
  \partial_z \,
  R_{L'}^{\mu\sigma}(\mathbf{r},\varepsilon),
\end{equation}
and $d_\text{at}$ is the atomic layer
thickness.\cite{2004-03//mavropoulos/papanikolaou/dederichs} The
transmission probability matrix $\mathbf{\Gamma}(\varepsilon)$ is
calculated between atomic pairs of the left and right lead, selected
in a way that the whole cross-section of the lead is covered (see
Fig.~\ref{fig:setup}). As it was shown previously, using one atomic
layer on each side yields well converged results in the fcc lattice
system.\cite{2004-03//mavropoulos/papanikolaou/dederichs} We verified
it for the TL system by varying the number of atomic layers used for
the transmission probability calculation. The difference between
$\mathbf{\Gamma}$ calculated using one and two layers in the leads did
not exceed 0.3\%.

While quantities within the KKR-GF approach are usually evaluated on a
complex energy contour, the transport coefficients should be
calculated from the transmission probability evaluated on the real
energy axis. In the proximity of the real energy axis, a linear
dependence of $\mathbf{\Gamma}(\varepsilon)$ on $\Im(\varepsilon)$ is
expected.\cite{2008-02//buruzs/szunyogh/weinberger} Thus, the
transmission probability $\mathbf{\Gamma}(\varepsilon)$ can be
calculated for several small values of the $\varepsilon$ imaginary
part [but setting $\Im(\varepsilon)$ large enough to ensure numerical
stability in the Green function calculation] and an estimate of
$\mathbf{\Gamma}(E)$ [where $E=\Re(\varepsilon)$] is then obtained as
a linear extrapolation of $\mathbf{\Gamma}(\varepsilon)$ to
$\Im(\varepsilon)=0$. We indeed found a very close to linear behavior
of $\mathbf{\Gamma}(\varepsilon)$ on small values of
$\Im(\varepsilon)$ and for all production runs, we calculated
$\mathbf{\Gamma}$ for two $\Im(\varepsilon)$ values corresponding to a
smearing temperature of 20~K and 10~K and extrapolated the results to
0~K.

The transport coefficients $\mathbf{L}_n$ were evaluated by a
numerical integration of $\mathbf{\Gamma}(\mathbf{k}_{\parallel},E)$
over a set of discrete values of the momentum $\mathbf{k}_{\parallel}$
and energy $E$ as
\begin{equation}\label{eq:Ln}
  \mathbf{L}_n=
  -\int \text{d}E \ \frac{\partial f_T(E)}{\partial E} \
  (E-E_\text{F})^n 
  \int_{\text{SBZ}} \! \text{d}\mathbf{k}_{\parallel}
  \mathbf{\Gamma}(\mathbf{k}_{\parallel},E).
\end{equation}
Here,
$f_T(E)=\left[\text{exp}\left(\frac{E-E_\text{F}}{k_\text{B}T}\right)+1\right]^{-1}$
is the Fermi-Dirac distribution function with $T$ corresponding to the
temperature of the MC simulation, $E_\text{F}$ is the Fermi energy and
$\text{d}\mathbf{k}_{\parallel}$ is the integration element in the
SBZ. For each temperature and system,
$\mathbf{\Gamma}(E)=\int_{\text{SBZ}}\text{d}\mathbf{k}_{\parallel}
\mathbf{\Gamma}(\mathbf{k}_{\parallel},E)$ was calculated on an
individual grid of 21 equidistant $E$ points in the range from $-10\
k_\text{B}T$ to $+10\ k_\text{B}T$ as the
$(E-E_\text{F})^n\partial{f_T}/\partial{E}$ factors become negligible
at $\pm 10\ k_\text{B}T$. We tested a four times denser $E$ grid for
the TL system and found no significant change in the results. Finally
we averaged over the non-collinear MC configurations at a given
temperature obtaining $\langle \mathbf{L}_n \rangle_T$.

The transport quantities, namely, electrical conductance $G$,
electrical resistance $R$, charge Seebeck coefficient $S_{\text{C}}$
and spin Seebeck coefficient $S_{\text{S}}$, are finally calculated
using the well-known formulas
\begin{align}\label{eq:G}
  G^{\sigma\sigma'}(T)&=\frac{e^2}{h}\langle {L}_0^{\sigma\sigma'}\rangle_T\\
  G&={\textstyle\sum_{\sigma\sigma'}G^{\sigma\sigma'}}\\
  R&=\frac{1}{G}\\
  S_\text{C}&=-\frac{\sum_{\sigma\sigma'}\langle L_1^{\sigma\sigma'}\rangle_T}
  {eT\sum_{\sigma\sigma'}\langle L_0^{\sigma\sigma'} \rangle_T}\\
  S_\text{S}&=-\frac{
    \langle L_1^{\uparrow\uparrow}\rangle_T+\langle L_1^{\downarrow\uparrow}\rangle_T
    -\langle L_1^{\downarrow\downarrow}\rangle_T-\langle L_1^{\uparrow\downarrow}\rangle_T
  }
  {eT\sum_{\sigma\sigma'}\langle L_0^{\sigma\sigma'}\rangle_T}
  .
\end{align}

\section{Results}\label{sec:res}

\subsection{Electronic structure of the ground state}\label{ssec:res-esgs}

\begin{figure}
  \includegraphics[width=\columnwidth]{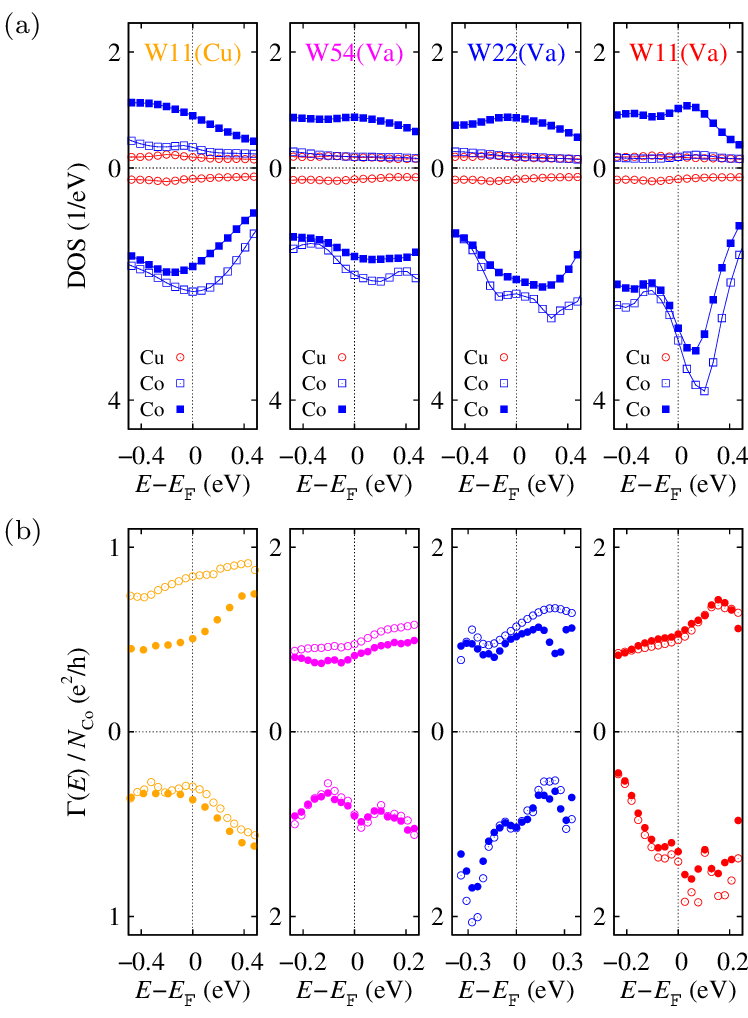}
  \caption{(Color online) (a) The electron density of states (DOS) for
    majority and minority spin channel (upper and reverse scale lower
    part of the graphs, respectively) averaged over four Cu layers
    adjacent to the scattering region (circles) and all Co atoms in
    the system (squares). Data corresponding to the spin-ordered and
    spin-disordered case (calculated at MC simulation $T\approx
    1100$~K) is shown as open and filled symbols, respectively. (b)
    Transmission probability $\Gamma$ divided by the number of Co
    atoms ($N_\text{Co}$) at the most narrow constriction of the
    respective system ($N_\text{Co}$ is 9, 4, 2, 1 for the TL, W54,
    W22, W11 systems, respectively), corresponding to the systems in
    (a). Open and filled symbols correspond to the spin-ordered and
    average spin-disordered case, respectively. In the spin-disordered
    case, upper and reverse scale lower part of the graphs depict
    $\Gamma^{\uparrow}=\Gamma^{\uparrow\uparrow}+(\Gamma^{\uparrow\downarrow}+\Gamma^{\downarrow\uparrow})/2$
    and
    $\Gamma^{\downarrow}=\Gamma^{\downarrow\downarrow}+(\Gamma^{\uparrow\downarrow}+\Gamma^{\downarrow\uparrow})/2$,
    respectively. The MC simulation temperature of the spin-disordered
    data (left to right) corresponds to 1100, 300, 400, and 300~K. The
    color coding of data sets in (b) is consistent with the system
    labels in Fig.~\ref{fig:geom}.}
  \label{fig:dostran}
\end{figure}

Before we proceed to the analysis of the spin disorder effect on the
transport properties, we want to point out the characteristic features
of the studied systems, as well as their similarities and/or
differences. For that purpose, the electron density of states (DOS) of
selected systems is depicted in Fig.~\ref{fig:dostran}(a). For
brevity, we refer to the majority and minority spin channel as
$\uparrow$ and $\downarrow$, respectively and we use the abbreviations
for the model systems as introduced in Fig.~\ref{fig:geom}. For all
systems in the ordered magnetic state, the DOS$^\uparrow(E_\text{F})$
of Co atoms (open squares) as well as the
DOS$^{\uparrow/\downarrow}(E_\text{F})$ of Cu atoms (open circles) is
rather low due to formally filled $d$ orbitals. In the TL system,
partially occupied minority spin $d$ orbitals of Co atoms yields the
DOS$^\downarrow(E_\text{F})$ about three times larger than the
DOS$^\uparrow(E_\text{F})$ (not shown). The one-dimensional character
of the nanowires is expected to manifest itself via Van Hove
singularity features in the DOS. The DOS$^\downarrow$ at around
$E_\text{F}$ is indeed raised for all nanowires, but a pronounced peak
can be seen only in the W11(Va) system. We note that a strong
asymmetry of the DOS around $E_\text{F}$, as seen for W11(Va), was
suggested as an indicator of a large Seebeck
coefficient.\cite{2011-01//vu/sato/katayama-yoshida}

In Fig.~\ref{fig:dostran}(b), we show the transmission probability
$\Gamma$ as a function of the energy $E$ around the Fermi level,
divided by the number of Co atoms ($N_\text{Co}$) at the most narrow
constriction of the respective system ($N_\text{Co}$ is 9, 4, 2, 1 for
the TL, W54, W22, W11 systems, respectively). The transmission
probability $\Gamma^\uparrow$ at $E_\text{F}$ is in the spin-ordered
state (open symbols) slightly smaller than 1 and a slowly growing
function of $E$ consistently for all systems. The $\Gamma^\downarrow$
exhibits much richer variation of its character. For example, despite
the similarity of the DOS between W54(Va) and W22(Va), the
corresponding $\Gamma^\downarrow$ as a function of $E$ is very
different. A discussion of the spin disorder effect on the DOS and
$\mathbf{\Gamma}$ will follow later, together with an analysis of
individual transport properties.

\subsection{Spin disorder}\label{ssec:res-sd}

\begin{figure}
  \includegraphics[width=\columnwidth]{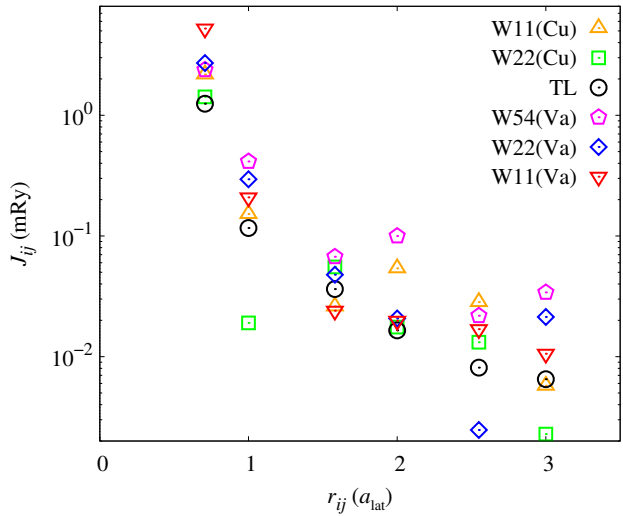}
  \caption{(Color online) Absolute value of the leading exchange
    coupling parameters $J_{ij}$ in logarithmic scale as a function of
    the Co inter-site distance $r_{ij}$. The color coding of data sets
    is consistent with the system labels in Fig.~\ref{fig:geom}.}
  \label{fig:jij}
\end{figure}

As outlined in the previous section, the spin disorder is determined
by the fluctuations of the magnetic moments obtained from snapshots of
MC simulations of the classical Heisenberg Hamiltonian. The exchange
coupling parameters $J_{ij}$ between the Co atoms were calculated for
all pairs for which the distance $r_{ij}\leq 3\,a_{\text{lat}}$. We
verified that a further increase of $r_{ij}$ did not affect
results. The $J_{ij}$ parameters between the periodic images of the
nanowires due to the in-plane periodicity were neglected, as their
values were at least 3 orders of magnitude smaller than the nearest
neighbor exchange coupling. Figure~\ref{fig:jij} shows the absolute
value of the leading $J_{ij}$ terms in logarithmic scale, showing that
the nearest and next-nearest neighbor interactions are dominant.

\begin{figure*}
  \includegraphics[width=\textwidth]{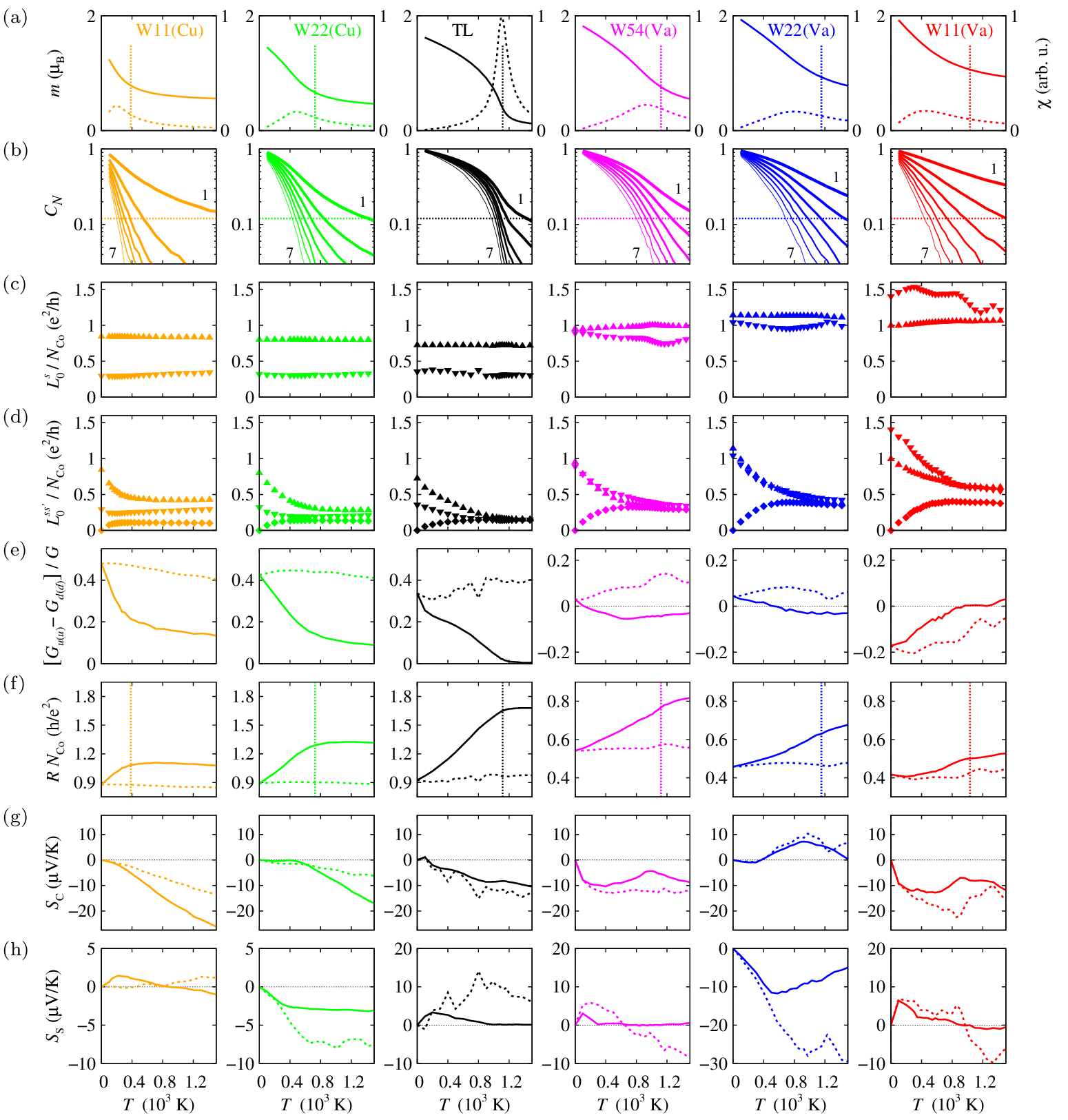}
  \caption{(Color online) (a) Monte Carlo (MC) site averages of the Co
    atoms magnetic moment $m$ (solid line) and magnetic susceptibility
    $\chi$ (dashed line). (b) Spatial correlation $C_N$ of the
    magnetic moment orientation between $N$-th nearest neighbor layers
    in the $z$ direction. Thick-to-thin line corresponds to
    $1^\text{st}$ to $7^\text{th}$ nearest neighbors,
    respectively. Vertical dotted line in (a) indicates temperature
    for which $C_3$ drops under 0.12 [dotted line in (b)]. (c)
    Electrical conductance of the ordered magnetic configuration, the
    temperature dependence enters via Fermi function
    smearing. Triangles pointing up/down correspond to
    $\uparrow$/$\downarrow$ spin, respectively. (d) Electrical
    conductance of the spin-disordered magnetic configurations
    calculated by MC method. Triangles pointing up/down and diamonds
    correspond to $\uparrow\uparrow$/$\downarrow\downarrow$ and
    average of $\uparrow\downarrow$ and $\downarrow\uparrow$ of spin
    matrix elements, respectively. (e) Polarization of the electrical
    conductance. (f) Total electrical resistance. Position of the
    vertical dotted line is equivalent to (a). (g) Charge and (h) spin
    Seebeck coefficient. In (e-h), dashed and solid lines correspond
    to the spin-ordered and -disordered data, respectively. The color
    coding of data sets in the individual columns corresponds to the
    system labels in Fig.~\ref{fig:geom}.}
  \label{fig:results}
\end{figure*}

The magnetization and susceptibility as a function of the MC
simulation temperature is shown in Fig.~\ref{fig:results}(a). Since
there is no real critical temperature in nanostructures, we define
$T_\text{c}$ as a crossover temperature at the magnetic susceptibility
peak. While the TL system exhibits critical-like behavior at around
1100~K (sharp drop of the magnetic moment and divergence character of
the magnetic susceptibility), all nanowires retain a rather large
magnetic moment until very high temperatures, resembling macro-spin
character. The reduction of magnetic moment at high temperatures is
less pronounced for the nanowires with smaller cross-section.  The
divergence character of their magnetic susceptibility, as a possible
indication of the crossover temperature, is strongly suppressed and
the corresponding peaks are shifted to much lower temperatures.

As we will show later, certain features of the spin disorder effects
on the transport properties could not be correlated with the position
of the susceptibility peak. Therefore, we explore an alternative sign
of the magnetic order loss, the spatial correlation $C_N(T)$ as
defined in Eq.~\eqref{eq:zcorr}.  In Fig.~\ref{fig:results}(b), the
$C_N(T)$ calculated between the moments of the $N$-th nearest neighbor
layers in the $z$ direction, is shown in logarithmic scale for
$N=\{1,\ldots,7\}$. The nearest neighbor spatial correlation $C_1$
(thickest line) when seen in linear scale (not shown) has a very
similar evolution with $T$ as the magnetic moment. The spatial
correlation between farther neighbors of the different systems
exhibits very diverse trends. In the case of TL, all farther-neighbor
$C_{N\geq 2}$ terms tightly follow the $C_1$ up to $T_\text{c}$. Above
this temperature, the falloff of $C_1$ and $C_2$ is less steep than
that of $C_{N\geq 3}$. This is consistent with the presence of
short-range magnetic order well above $T_\text{c}$ while the
long-range order is quickly suppressed. In the case of nanowires, the
$C_N$ tend to quickly deviate from each other already at low
temperatures. The nanowires embedded in Cu share a similar picture
with the TL, exhibiting a fast decay of $C_{N\geq 3}$. The nanowires
embedded in vacuum display a somewhat different trend, with no clear
separation in the falloff of $C_1$ and $C_2$ with respect to $C_{N\geq
3}$. Instructive is a comparison of the W22(Cu) and W11(Va) nanowires,
for which the susceptibility peaks at around the same temperature. The
$C_1$ and $C_2$, however, stay quite large at high $T$ in the case of
W11(Va), suggesting well preserved short-range magnetic order and
possibly different character of the spin disorder effect on the
transport properties in comparison with W22(Cu). Using the spatial
correlation, we determine an independent measure of the long-range
magnetic order loss.  Especially for the TL, the crossover point
($T_\text{c}\approx 1100$~K) is very well approximated by using the
condition that the $C_3$ value falls below 0.12 [indicated as dotted
vertical and horizontal line in Fig.~\ref{fig:results}(a) and
\ref{fig:results}(b), respectively]. However, this is not the case for
the nanowire systems, where the difference between the $T_\text{c}$
and a crossover point determined from the falloff of $C_3$ is quite
large. This suggests that the loss of order in the nanowires is more
gradual, resulting in a strong drop of $C_3$ at higher temperatures
than $T_\text{c}$.

\subsection{Transport properties}\label{ssec:res-tran}

The electrical conductance of the spin-ordered and spin-disordered
states as a function of temperature is shown in
Fig.~\ref{fig:results}(c) and Fig.~\ref{fig:results}(d),
respectively. As it can be seen in Fig.~\ref{fig:results}(c), the
temperature dependence of the electrical conductance calculated using
only the Fermi function smearing for the ordered magnetic
configuration is very weak. Interestingly, the conductance of both
W11(Cu) and W22(Cu) is very similar to the TL system, although the
number of Co atoms in the supercells containing a nanowire is much
lower. On the contrary, the relative magnitude of the conductance via
the majority and minority spin channels is reversing with decreasing
thickness of the nanowires embedded in vacuum. As expected, the
temperature effect on the electrical conductance via disordered
magnetic configurations is significant
[Fig.~\ref{fig:results}(d)]. The individual spin matrix elements of
the conductance tend to converge at high temperatures.  While the
convergence is rather quick for the TL, it is suppressed as the
cross-section of the nanowires decreases. This correlates with the
fact that the $\uparrow$ and $\downarrow$ channels of the DOS and
transmission probability for the nanowires are not equalized by the
spin disorder even at $T\approx 1100$~K [shown as filled symbols in
Fig.~\ref{fig:dostran}(a) and \ref{fig:dostran}(b) - only
W11(Cu)]. The resulting electrical conductance polarization
[Fig.~\ref{fig:results}(e)] shows that small variations in the
geometry of nanowires can lead to large differences in the
polarization [e.g., negative sign for W11(Va)]. Furthermore, the spin
disorder has, in general, indeed a significant influence (solid
line). Besides strong suppression of the polarization [W11(Cu),
W22(Cu), TL], it can lead to sign reversal (nanowires embedded in
vacuum).

The well-known effect of spin disorder resistivity can be seen in
Fig.~\ref{fig:results}(f). A characteristic kink in the electrical
resistance of the TL correlates with the crossover point observed in
the MC data. This kink is present also in the nanowires embedded in
copper (although it is not as sharp) and the resistance saturates
above a certain temperature. The position of the kink (indicated by a
vertical dotted line) correlates rather with the loss of long-range
magnetic order ($C_{N\geq 3}\lessapprox 0.12$) than with the peak in
magnetic susceptibility $\chi$. The kink is virtually missing in the
nanowires embedded in vacuum where the resistance grows throughout the
whole considered temperature range. Yet, a small change of slope in
$R(T)$ can be identified, which again correlates rather with the
long-range magnetic order loss than with the susceptibility peak.

While features of electrical conductance can be, for some simple
systems, related to the electronic density of states, it is next to
impossible to find such relation in case of the Seebeck coefficient,
except perhaps in the low-temperature limit of some model systems. The
reason is that the Seebeck coefficient is a product of two transport
coefficients $1/L_0$ and $L_1$, where the former may already have not
much relation to the DOS and the latter is resulting from
contributions of the transmission probability $\mathbf{\Gamma}(E)$,
with maximum weight at $\lvert{E-E_\text{F}}\rvert\approx 1.5\
k_\text{B}T$ and significant weight up to as far as
$\lvert{E-E_\text{F}}\rvert\approx 5\ k_\text{B}T$.

In Fig.~\ref{fig:results}(g) and Fig.~\ref{fig:results}(h), we present
the results of the conventional charge ($S_\text{C}$) and spin
($S_\text{S}$) Seebeck coefficient, respectively. The influence of
temperature is obviously non-trivial, even if only the effect due to
the Fermi function smearing for the ordered spin configuration is
considered (dashed line). Looking first at $S_\text{C}$, a very fast
onset can be seen for W54(Va) and W11(Va), quickly reaching almost its
maximum value already at around or even well below room
temperature. This result could be interpreted in the light of the
hypothesis that a large Seebeck coefficient can be predicted from a
steep DOS$(E)$ slope at $E_\text{F}$ due to Van Hove singularity in
quantum wires. While this is consistent for the W11(Va) nanowire, the
DOS of the W54(Va) nanowire is very flat at around $E_\text{F}$. The
explanation lies in the character of the transmission probability
[Fig.~\ref{fig:dostran}(b)]. For both systems and both spin channels
the slope of $\Gamma(E)$ at around $E_\text{F}$ is positive and
relatively large leading to a fast growth of $S_\text{C}(T)$. The
resulting difference of the $\uparrow$ and $\downarrow$ slope of
$\Gamma(E_\text{F})$ yields somewhat smaller $S_\text{S}$. A
complementary argument showing the inability of the DOS to reflect the
Seebeck coefficient comes from a comparison of the W54(Va) and W22(Va)
nanowires. Despite a close similarity in the DOS
[Fig.~\ref{fig:dostran}(a)], the effective slope of
$\Gamma^\downarrow(E)$ at around $E_\text{F}$ is opposite in sign for
these two systems [Fig.~\ref{fig:dostran}(b)], which leads to a
negligible $S_\text{C}$ and an enhanced $S_\text{S}$ at room
temperature for the W22(Va) nanowire
[Fig.~\ref{fig:results}(g-h)]. The growth of $S_\text{S}(T)$ slows
down at higher $T$ due to sharp kinks in $\Gamma^\downarrow$ at
approximately -0.3~eV and 0.2~eV [Fig.~\ref{fig:dostran}(b)] while the
slope difference of $\Gamma^\uparrow$ and $\Gamma^\downarrow$ at
elevated temperatures leads to a sign reversal of $S_\text{C}$ in
comparison with all other model systems.

The effect of spin disorder on the Seebeck coefficients [solid line in
Fig.~\ref{fig:results}(g-h)] is again generally quite pronounced as it
was in the case of the electrical conductance, its polarization and
the electrical resistance. An interesting observation is the rather
strong enhancement of $S_\text{C}$ due to the spin disorder for the
nanowires embedded in copper. The origin is obvious when looking at
the transmission probability of the W11(Cu) nanowire in
Fig.~\ref{fig:dostran}(b). The spin disorder (shown at $\approx
1100$~K) causes a drop of the majority spin $L_0$ coefficient while
the kink at $\Gamma^\uparrow$ very close to $E_\text{F}$ leads to a
rise in the majority spin $L_1$ coefficient. The resulting $L_1/L_0$
ratio is then significantly enhanced. Very similar behavior is seen
for the W22(Cu) nanowire, where the $S_\text{C}$ enhancement due to
the spin disorder is shifted to higher temperatures. This is caused by
the already mentioned kink at $\Gamma^\uparrow$ positioned slightly
away from $E_\text{F}$ (not shown). Furthermore, the spin disorder is
expected to significantly suppress the spin Seebeck coefficient which
is, in general, indeed observed. However, for the W22(Cu) and W22(Va)
nanowires, the $S_\text{S}$ remains relatively large even at very high
temperatures due to the non-vanishing difference of the $L_1$
transport coefficient for the majority and minority spin channels.

\section{Summary}\label{sec:sum}

We investigated the effect of temperature induced spin-disorder on the
transport through several Co nanostructures embedded between Cu
leads. The calculation of the transport properties confirmed that, at
elevated temperatures, spin disorder affects the value of the
transport coefficients both qualitatively and quantitatively, and
therefore cannot be neglected in a theoretical analysis.
Additionally, we find that there is no clear connection between the
transport properties and the density of states, due to the complex
convolution of the Fermi function derivative and the energy dependent
transport coefficients. These conclusions apply to the resistance,
charge and spin Seebeck coefficients.

The well-known spin-disorder contribution to the resistance is found
to be significant in the systems we studied. The temperature, at which
the characteristic kink in the resistance is observed, can be related
to the onset of the long-range magnetic order loss, determined from
the spatial correlation of the fluctuating local magnetic moments.  We
find a non-trivial behavior of the charge and spin Seebeck coefficient
as a function of temperature, that does not follow a clear universal
semi-quantitative or even qualitative rule, as a number of effects are
factored in for its calculation, including the fluctuations of local
magnetic moments, their temperature-dependent correlation, the quantum
confinement due to the nanostructure geometry, the participating
conducting states due to the Fermi distribution, and the interface
transmission. Furthermore, we showed that a decrease, an enhancement
or even a change of sign of the charge and spin Seebeck coefficients
can result from an interplay of the spin-disorder and the geometry in
a particular microscopic structure.

It is obvious that the spin disorder constitutes only one of many
effects that contribute to the spin-caloric transport at high
temperatures, others being phonons or magnon-assisted spin transport,
not considered in this work. However, our results show that
spin-disorder at high temperatures cannot be neglected for a
quantitative or even qualitative description of thermoelectric and
spin-caloric coefficients in magnetic nanostructures.

\acknowledgments

Support from the Deutsche Forschungsgemeinschaft (SPP 1538 ``Spin
Caloric Transport'') is gratefully acknowledged. Computational
resources were provided by the groups of M. Le\v{z}ai\'c, S.  Lounis
and Y. Mokrousov at the PGI-1 at Forschungszentrum J\"ulich as well as
from the JARA-HPC from RWTH Aachen University under project
jara0051. We are grateful to D. Comtesse, P. Entel, P. Kratzer,
V. Popescu, C.\,E. Quiroga and L. Szunyogh for numerous enlightening
discussions.

\bibliography{references}

\end{document}